\documentclass[aps,prb,twocolumn,showpacs,superscriptaddress,amsmath,amssymb,floatfix,10pt]{revtex4-1}  
\usepackage{graphicx}  
\usepackage{dcolumn}   
\usepackage{bm}        
\usepackage{amssymb}   
\usepackage{framed}
\usepackage{amsmath}
\usepackage{multirow}
\usepackage{hhline}
\usepackage{color}
\definecolor{bluegreen}{rgb}{0,0.2,0.8}
\usepackage{subfigure,amsmath,verbatim,moreverb}
\usepackage{tabularx}
\usepackage{adjustbox}
\usepackage{lipsum}

\begin{document}

\widetext
\title{Performance of Tao-Mo semilocal density functional in projector-augmented-wave method}

\author{Subrata Jana}
\email{subrata.jana@niser.ac.in}
\affiliation{School of Physical Sciences, National Institute of Science Education and Research, HBNI, 
Bhubaneswar 752050, India}
\author{Abhilash Patra}
\affiliation{School of Physical Sciences, National Institute of Science Education and Research, HBNI, 
Bhubaneswar 752050, India}
\author{Prasanjit Samal}
\email{psamal@niser.ac.in}
\affiliation{School of Physical Sciences, National Institute of Science Education and Research, HBNI, 
Bhubaneswar 752050, India}

\date{\today}

\begin{abstract}
We assess the performance of Tao-Mo semilocal exchange
correlation (TM) functional [J. Tao and Y. Mo, Phys. Rev. Lett. {\bf
117}, 073001 (2016)] using projector-augmented-wave method with the plane
wave basis set in Vienna {\it{ab initio}} simulation
package (VASP). The meta-GGA level semilocal functional TM is an all
purpose exchange-correlation functional which performs accurately for the
wide range of molecular and solid state properties. The exchange
functional part of TM is designed from the density matrix expansion (DME)
technique together with the slowly varying fourth order gradient
expansion. The correlation functional of the corresponding exchange is
based on Tao-Perdew-Staroverov-Scuseria (TPSS) functional. We assess the
performance of TM for solid state lattice constants, bulk moduli, band
gaps, cohesive energies and magnetic moments of solids. It has been
established that in plane wave basis the TM functional performs
accurately in predicting all the solid state properties in semilocal
level.

\end{abstract}

\maketitle

\section{introduction}
The calculations of the electronic structure of molecules and solids are
done mostly within the framework of Kohn-Sham (KS) density functional
theory~\cite{ks65,engelbook}. The accuracy of density functional theory
(DFT) depends upon the accuracy of the exchange-correlation (XC)
functionals which contain all the many electron effects. In principle
the exact form of XC functional is unknown. Therefore, one need to
approximate it from different physical perspective. The accuracy of the
XC functionals are classified through the Jacob's ladder~\cite{jacob},
where each rung of the ladder adds an extra ingredients. The first three
rungs of the Jacob's ladder are classified as local density approximation
(LDA)~\cite{lda}, the generalized gradient approximations
(GGAs)~\cite{PW86,B88,LYP88,PW91,B3PW91,PBE96,AE05,ZWu06,PBEsol,con4,
con1,con3} and meta-generalized gradient approximations
(meta-GGAs)~\cite{BR89,VSXC98,MO6L,TPSS03,revTPSS,con2,SCAN15,
Tao-Mo16}. The LDA, GGAs and meta-GGAs utilize density, gradient of
density and Kohn-Sham kinetic energy density as its main ingredient,
therefore all are semilocal in nature. The XC functionals based on the
semilocal quantities are very attractive because of low computational
cost and high accuracy. The semilocal functionals are very accurate in
describing  several  thermochemical
properties~\cite{CJCramer09,Quest12,Yangreview,Becke14,VNS03,PHao13,
LGoerigk10,LGoerigk11,tmpccp,phmk}, bond lengths~\cite{VNS03},
equilibrium  lattice constants~\cite{PBlaha09,FTran16,YMo16,YMoCPL,phmk,mapkw,smrkp,
mgga-vasp}, bulk modulus, cohesive
energy~\cite{PBlaha09,FTran16,YMo16,YMoCPL,phmk,mapkw,smrkp,mgga-vasp}
and  solid  state  surface
properties~\cite{PBlaha09,FTran16,YMo16,YMoCPL,mapkw,smrkp,mgga-vasp}.  
Several accurate semilocal density functionals have been developed for last
couple of decades from different new perspective.
The semilocal functionals are developed from exchange
hole~\cite{Tao-Mo16} or by satisfying exact
constraints~\cite{PBE96,SCAN15}. Functionals which are designed by
satisfying exact constraints non-empirical in nature and very popular
in quantum chemistry and solid state
physics~\cite{PBE96,SCAN15,Tao-Mo16}. Beyond these classes of XC
functionals highly empirical XC
functionals~\cite{MO6L} are also proposed by properly parameterize with
a test set. Those functionals are very accurate in describing
thermochemical properties~\cite{Quest12} but not popular for solid state
systems. Therefore, it is always interesting to test the robustness of a 
functional which is universal in nature i.e., useful 
both for quantum chemistry and solid state systems.

Recently, Tao-Mo (TM)~\cite{Tao-Mo16} developed an accurate semilocal functional
based on the density matrix expansion (DME). The slowly varying forth order gradient
correction of exchange enhancement factor is introduced within DME by
interpolating it with DME enhancement factor. The DME based
enhancement factor is accurate for localized electron systems, whereas,
slowly varying density correction is accurate for solid state
systems. The correlation part TM functional has been derived
from one electron self-interaction free Tao-Perdew-Staroverov-Scuseria
(TPSS) functional~\cite{TPSS03}. The TM functional is very accurate 
in describing both the thermochemical~\cite{tmpccp} and solid state properties 
in semilocal level~\cite{YMo16,YMoCPL}. TM functional is extensively tested in
all electron code but has not been benchmarked in plane wave basis set. Due
to its higher degree of accuracy for solid state properties it is always
interesting to test the TM functional using plane wave basis set.  
Present paper assess the performance of TM functional using projector-augmented-wave
(PAW)~\cite{paw1,paw2} method with plane wave basis set in Vienna {\it{ab
initio}} simulation package (VASP)~\cite{vasp1,vasp2,vasp3}. For the
benchmark calculation  of TM functional we choose solid state lattice
constants, bulk moduli, semiconductor band gaps, cohesive energies and
magnetic moments of ferromagnetic materials. We compare the
performance of TM functional against other popular semilocal functionals
like local spin density approximation (LSDA)~\cite{lda},
Perdew-Burke-Ernzerhof (PBE)~\cite{PBE96} functional and its revised form
for solid state system (PBEsol)~\cite{PBEsol}, meta-GGA level
TPSS~\cite{TPSS03}, revTPSS~\cite{revTPSS} and recently proposed strongly
constrained and appropriately normed (SCAN)~\cite{SCAN15} functional. It
is shown that the performance of TM functional
is very accurate in describing lattice constants, bulk
moduli, cohesive energies and magnetic properties except band gap.
In describing the band gap, SCAN meta-GGA performs in more satisfactory
way. To put our comparison in a broader perspective we also include
Heyd-Scuseria-Ernzerhof (HSE)~\cite{HSE03,mpsk,jmhk,shk,bgref} hybrid
functional for studying magnetic properties.

Present paper is organized as follows. In the following we will discuss
the details of the implementation of TM based semilocal functionals in
VASP. Next we do the benchmark performance of the TM based
functionals with other popular GGA and meta-GGA level semilocal
functionals for lattice constants, bulk moduli, band gaps, cohesive
energies and magnetic moments of ferromagnetic materials.

\section{Theoretical Background}
  The PAW~\cite{paw1,paw2} method was first proposed by Bl\"{o}chl~\cite{paw1} and later
  it is adopted by Kresse and Joubert~\cite{paw2}. The PAW potential 
reserves both the computational efficiency and accuracy of ultrasoft pseudo-potential
(UPP)~\cite{uspp} and all electron (full-)potential. The accuracy of PAW is same as the
full potential linearized augmented plane wave (plus local potential) implemented in other
codes~\cite{wien2k}. In PAW method, any semilocal operator (i.e.,kinetic energy operator
or density operator) is presented by plane wave (PW) expansion for valance region. The
core region is presented by projecting it on the radial grid at the atom center.
Therefore, the core region is well presented in this method and its accuracy is same as
the all electron (AE) calculation. After doing all these the additional part is subtracted
from the additive augmentation of core and valence. The general 
implementation of all the meta-GGA functional in VASP is based on the method proposed
by Sun et. al.~\cite{mgga-vasp}. In this method the KS kinetic energy density also divided
in core-valance region. We implemented the TM semilocal functional by locally modifying
the meta-GGA routine implemented in VASP code. The self-consistence
exchange-correlation potential computed for meta-GGA in VASP is based on generalized KS
(gKS) framework. In gKS, the exchange-correlation potential for meta-GGA is defined as,
\begin{eqnarray}
v_{xc}\Psi_{i} &=& \Big[\frac{\partial(\rho\epsilon_{xc})}{\partial\rho}
-\vec{\nabla}\frac{\partial(\rho\epsilon_{xc})}{\partial\vec{\nabla}\rho}\Big]
\Psi_{i}-\frac{1}{2}\vec{\nabla}\Big(\frac{\partial(\rho\epsilon_{xc})}{\partial\tau}\Big)
\vec{\nabla}\Psi_{i}\nonumber\\
&-&\frac{1}{2}\frac{\partial(\rho\epsilon_{xc})}{\partial\tau}\vec\nabla^2\Psi_{i}~.
\label{eq1}
\end{eqnarray}
Therefore, in addition to the partial derivative of XC functional with respect to density and 
gradient of density, one need to perform the partial derivative with respect to KS kinetic 
energy density. Thus, for meta-GGA one need to calculate the partial
derivative of XC functional with respect to density, gradient of density and KS kinetic energy
density. In terms of enhancement factor the general formulation of the exchange functional
in meta-GGA level is described as,
\begin{equation}
E_{x}=-\int~d\mathbf{r}~\rho(\mathbf{r})\epsilon_{x}^{unif}F_{x}^{meta-GGA}[\rho,\nabla\rho,\tau]~.  
\label{eq2}
\end{equation}
In the present case of TM functional the enhancement factor becomes,
\begin{equation}
F_{x}^{meta-GGA}[\rho,\nabla\rho,\tau]=F_x^{TM}~,
\end{equation}
which have the following semilocal form,
\begin{equation}
F_x^{TM} = wF_x^{DME} + (1-w)F_x^{sc},
\label{eq3}
\end{equation}
where $F_x^{DME}=1/f^2+7R/(9f^4)$ is the density matrix expansion based enhancement
factor~\cite{Tao-Mo16} (with $R=1+595(2\lambda - 1)^2p/54 -[\tau -(3\lambda^2-\lambda+1/2)(\tau-\tau^{unif}-|\nabla\rho|^2/(72\rho))]/\tau^{unif}$)
and   $F_x^{sc}=\Big[1+10\Big\{\Big(\frac{10}{81}+\frac{50p}{729}\Big)p+
 \frac{146}{2025}\tilde{q}^2-\Big(\frac{73\tilde{q}}{405}\Big)\Big[\frac{3\tau^w}
 {5\tau}\Big](1-\frac{\tau^w}{\tau}\Big)\Big\}\Big]^{\frac{1}{10}}$ is the slowly varying
 forth order gradient expansion to the TM semilocal functional~\cite{Tao-Mo16}.
 In TM functional $w$ is used as the weight factor between DME
 expansion and slowly varying density correction, which is the functional of meta-GGA
 ingredient $z=\tau_W/\tau$ (where $\tau_W$ is the von Weizs\"{a}cker kinetic energy
 density). For the details of the mathematical expression and parameter related to the TM
 functional readers are suggested to go through the refs~\cite{Tao-Mo16,YMo16}. It is
 noteworthy to mention that the discontinuity effect and  convergence issue of the terms
 related to the $z$ in TM functional is removed as it was suggested by Sun et.
 al.~\cite{mgga-vasp}. In meta-GGA subroutine of the VASP code we calculate analytically
 the terms related to the partial derivatives i.e.,
 $\frac{\partial(\rho\epsilon_{x})}{\partial\rho}$, $
 \frac{\partial(\rho\epsilon_{x})}{\partial\vec{\nabla}\rho}$ and
 $\frac{\partial(\rho\epsilon_{x})}{\partial\tau}$. This completes the implementation of
 exchange part of TM functional. The correlation of TM functional is based on the
 Tao-Perdew-Staroverov-Scuseria (TPSS)~\cite{TPSS03} correlation. TM modifies 
the correlation part to be used for slowly varying density correction. This leads to the
two functional $-$ TM-TPSS (which uses TM exchange plus TPSS correlation) and TM (which
uses TM exchange plus modified TPSS correlation). Here we assess the performance of both
TM-TPSS and TM for all our solid state calculations. The spin density scaling 
relation is used in the VASP implementation of TM functional. To test the accuracy of 
all the functionals under study we calculate mean (relative) error (ME/MRE), mean absolute
(relative) error (MAE/MARE)  and the standard deviation of the (relative) error
(STDE/STDRE) which are defined as, ME = $\frac{1}{N}\sum_{i=1}^N(Y_i-y_i)$,
MAE = $\frac{1}{N}\sum_{i=1}^N|Y_i-y_i|$, STDE
=$\Big[\sum_i^N(Y_i-y_i)-\frac{1}{N}\sum_i^N(Y_i-y_i)\Big]^2$,
MRE=$\sum_i^N(Y_i-y_i)/y_i$, MARE=$\sum_i^N|Y_i-y_i|/|y_i|$ and
STDRE=$\Big[\sum_i^N(Y_i/y_i)-\frac{1}{N}(Y_i/y_i)\Big]^2$, where $Y_i$ and $y_i$ are the
calculated and experimental values respectively.

\section{Results and Discussions}

\subsection{Lattice constants}
\begin{table*}
\label{latt-cons}
\caption{Equilibrium lattice constant $a_0$ (in \AA)~ of
different solid structures. The mean error, mean absolute error, mean relative error, and
mean absolute relative error are reported in the last row are determined with respect to
the  ZPAE uncorrected experimental values. All the experimental reference values are
collected from ref.~\cite{bgref,YMo16}. The structures we consider here are A1 $=$
face-centered cubic, A2 $=$ diamond, A3 $=$ body-centered cubic, B3 $=$ zinc blende, and
B1 $=$ rock salt.    }
\begin{tabular}{c  c  c  c  c  c   c   c  c c}
\hline\hline
Solids&LSDA&PBE&PBEsol&TPSS&revTPSS&SCAN&TM-TPSS&TM&Expt.\\
\hline 
C (A2)&3.536 &3.573 &3.557 &3.572 &3.563 &3.555 &3.560 &3.554 & 3.567 \\
Si (A2)&5.400 &5.467 &5.433 &5.450 &5.436 &5.425 &5.423 &5.411 & 5.430 \\
Ge (A2)&5.648 &5.785 &5.704 &5.754 &5.710&5.687 &5.691 &5.672 &5.652 \\
SiC (B3)&4.332 &4.379 &4.359 &4.365&4.357 &4.352 &4.351 &4.344 &4.358 \\
BN (B3)&3.583 &3.625 &3.607 &3.624 &3.618&3.605 &3.615 &3.608 &3.607 \\
BP (B3)&4.490 &4.546 &4.521 &4.545 &4.531&4.521 &4.522 &4.510 & 4.538 \\
BAs (B3)&4.742 &4.817 &4.778 &4.810&4.787 &4.779 &4.775 &4.763 &4.777 \\
BSb (B3)&5.198 &5.280 &5.234 &5.270&5.242 &5.257 &5.227 &5.212 &n/a \\
AlP (B3)&5.433 &5.504 &5.470 &5.489&5.480 &5.478 &5.463 &5.450 &5.460 \\
AlAs (B3)&5.637 &5.732 &5.681 &5.707&5.685 &5.670 &5.669 &5.656 &5.658 \\
AlSb (B3)&6.120 &6.232 &6.168 &6.208 &6.180&6.173 &6.161 &6.143 &6.136 \\
$\beta-$GaN (B3)&4.503 &4.588&4.547 & 4.581&4.569 &4.524 &4.559 &4.549 &4.531 \\
GaP (B3)&5.425 &5.533 &5.474 &5.523 &5.499 &5.457&5.482 &5.464 &5.448 \\
GaAs (B3)&5.627 &5.763 &5.684 &5.737 &5.699&5.664 &5.681 &5.664 &5.648 \\
GaSb (B3)&6.067 &6.226 &6.130& 6.190&6.144 &6.117 &6.126 &6.102 &6.096 \\
InP (B3)&5.878 &6.001 &5.932 &5.989&5.965 &5.938 &5.945 &5.923 &5.866 \\
InAs (B3)&6.061 &6.211 &6.122&6.182 &6.144 &6.122 &6.126 &6.104 &6.054 \\
InSb (B3)&6.472 &6.651 &6.543&6.611 &6.565 &6.545 &6.546 &6.521 &6.479 \\
ZnS (B3)&5.403 &5.440 &5.355 &5.401&5.358 &5.370 &5.388 &5.364 &5.409 \\
ZnSe (B3)&5.570 &5.734 &5.634& 5.681& 5.625&5.652 &5.658 &5.633 &5.668 \\
ZnTe (B3)&5.995 &6.178 &6.064 &6.115&6.048 &6.077 &6.082 &6.056 &6.089 \\
CdS (B3)&5.758 &5.926 &5.824 &5.933&5.926 &5.856 &5.889 &5.857 &5.818 \\
CdSe (B3)&6.009 &6.195 &6.080 &6.192& 6.195&6.100 &6.133 &6.102 &6.052 \\
CdTe (B3)&6.405 &6.610 &6.291 &6.604& 6.610&6.521 &6.532 &6.497 &6.480 \\
MgO (B1)&4.145 &4.242 &4.206 & 4.224&4.222&4.184 &4.209 &4.202 &4.207 \\
MgS (B3)&5.580 &5.684 &5.642 &5.681 &5.673&5.634 &5.643 &5.629 &5.202 \\
MgSe (B1)&5.382 &5.501 &5.445 &5.491&5.476 &5.454 &5.456 &5.435 &5.400 \\
MgTe (B3)&6.365 &6.506 &6.439 &6.500&6.478 &6.452 &6.444 &6.422 &6.420 \\
CaS (B1)&5.570 &5.710 &5.632 &5.698 &5.694&5.683 &5.681 &5.657 &5.689 \\
CaSe (B1)&5.798 &5.955 &5.869 &5.947&5.932 &5.921 &5.919 &5.894 &5.916 \\
CaTe (B1)&6.215 &6.389 &6.291 &6.386&6.366 &6.375 &6.350 &6.317 &6.348 \\
SrS (B1)&5.910 &6.056 &5.973 &6.047 &6.040&6.031 &6.035 &6.007 &5.990 \\
SrSe (B1)&6.129 &6.297&6.203 &6.286 &6.270 &6.264 &6.264 &6.234 &6.234 \\
SrTe (B1)&6.531 &6.714&6.609 &6.708 &6.685 &6.693 &6.677 &6.641 &6.640 \\
BaS (B1)&6.289 &6.433 &6.362& 6.448&6.440 &6.441 &6.423 &6.390 &6.389 \\
BaSe (B1)&6.510 &6.681&6.577 & 6.670&6.657 &6.659 &6.659 &6.622 &6.595 \\
BaTe (B1)& 6.890&7.080&6.964 & 7.075&7.054 &7.071 &7.056 &7.012 &7.007 \\
Ag (A1)&4.001 &4.148 &4.052& 4.092&4.059 &4.084 &4.082 &4.067 &4.069 \\
Al (A1)&3.987 &4.043 &4.081& 4.014&4.009 &4.009 &3.984 &3.982 &4.032 \\
Cu (A1)&3.520 &3.634 &3.566 &3.568&3.538 &3.555 &3.528 &3.528 & 3.603 \\
Pd (A1)&3.844 &3.949 &3.878 &3.912&3.884 &3.906 &3.908 &3.894 &3.881 \\
K (A3)&5.029 &5.300 &5.222 &5.394& 5.349&5.262 &5.186 &5.167 &5.225 \\
Li (A3)&3.368 &3.441 &3.444 &3.458& 3.452&3.474 &3.400 &3.402 &3.477 \\
LiCl (B1)&4.977 &5.148 &5.071&5.123 &5.104 &5.097 &5.071 &5.047 &5.106 \\
LiF (B1)&3.940 &4.059 &4.006 &4.022&4.005 & 3.975&3.974 &3.969 &4.010 \\
NaCl (B1)&5.432 &5.648&5.558 &5.648& 5.616&5.526 &5.415 &5.496 &5.595      \\
NaF (B1)&4.437 &4.621&4.548 &4.599&4.569 & 4.475&4.498 &4.492 &4.609      \\
\hline\hline
ME(\AA)&-0.055&0.076&0.002&0.061&0.039&0.020&0.019&0.000&$-$\\
MAE(\AA)&0.072&{\bf\color{red}0.078}&0.041&0.065&0.053&0.041&0.045&{\bf \color{blue} 0.038}&$-$\\
\hline
STDE(\AA)&0.082&0.076&0.079&0.079&0.081&0.074&0.077&0.075&$-$\\
MRE(\%)&-1.045&1.361&0.027&1.059&0.657&0.310&0.274&-0.065&$-$\\
MARE(\%)&1.375&1.406&0.738&1.157&0.961&0.753&0.854&0.753&$-$\\
STDRE(\%)&1.573&1.407&1.469&1.483&1.537&1.423&1.524&1.478&$-$\\
\hline\hline
\end{tabular} 
\end{table*}

\begin{table*}
\label{latt-cons}
\caption{Bulk moduli, B$_0$ (GPa) of 20 solids are shown for different functionals. The LSDA, PBE, PBEsol,
TPSS, revTPSS results for Ag, Al, C, GaAs, Ge, Li, LiCl, LiF, NaF, Pd, Rh are collected
from ref.~\cite{mgga-vasp}. For rest of the solids and functionals we perform self
consistence calculations followed by the fitting of E vs V curve using Murnaghan equation
of states~\cite{eos} sing VASPKIT~\cite{vaspkit} post-processing code. The 
structures consider here are the same as it is given in Table-I. All the experimental
values are collected from ref.~\cite{YMo16,jmhk}.}
\begin{tabular}{c  c  c  c  c  c   c   c  c c}
\hline\hline
Solids&LSDA&PBE&PBEsol&TPSS&revTPSS&SCAN&TM-TPSS&TM&Expt.\\
\hline
Ag&138.5&90.9&118.9&110.0& 120.5&111.5&111.3&115.1&109.0\\
Al&83.7&77.3& 81.9& 85.6&85.7&85.7&91.7&93.2&79.4\\
AlAs&74.5&67.4&71.6&70.3&72.2&75.5&75.7&76.4&82.0\\
AlP&89.0&82.0&85.9&84.9&86.1&90.7&89.4&90.7&86.0\\
BP&168.0&156.2&162.5&155.7&158.3&166.5&164.4&165.8&173.0\\
C&465.8&433.2&450.2&430.3& 439.5&461.4&447.9&455.2&443.0\\
Cu&185.4&139.4&165.4&158.6&172.4&161.7&170.1&173.2&142.0\\
GaAs& 75.1&60.5& 69.9&64.8& 66.8&73.3&71.7&74.1&75.6\\
GaN&209.8&183.5&197.1&188.9&191.2&210.2&197.3&200.5&190.0\\
GaP&90.7&78.0&85.3&79.6&82.5&90.7&87.5&90.1&88.0\\
Ge& 70.5&59.4& 65.8&60.2& 65.0&71.4&58.4&58.4&75.8\\
K&4.6&3.7&3.8&3.4&3.4&3.7&4.0&4.0&3.7\\
Li&15.1&13.8&13.7&13.3&13.4&13.2&14.6&14.6&13.0\\
LiCl&41.5& 31.7&35.4& 33.4&34.0&35.8&36.2&36.7&35.4\\
LiF& 86.7&66.9& 72.2& 66.2&  68.9&81.2&79.6&80.2&69.8\\
NaF& 61.5&45.2&48.8& 42.9& 44.0&61.9&59.5&59.9&51.4\\
Pd&226.3& 169.4&205.2& 195.4& 209.7&194.8&191.3&199.5&195.0\\
Rh&315.6&256.4&295.0& 281.9&296.1&290.9&283.4&291.2&269.0\\
Si&95.4&87.9&92.7&91.3&93.0&98.7&96.3&98.4&99.2\\
SiC&221.5&205.1&213.7&217.7&221.3&225.7&217.8&221.1&225.0\\
\hline\hline
ME (GPa)&10.695&-9.870&1.335&-3.545&0.935&0.485&2.145&4.650&$-$\\
MAE (GPa)&{\bf\color{red}13.255}&9.950&7.615&7.285&8.235&7.785&{\bf\color{blue}7.195}&8.290&$-$\\
\hline
STDE (GPa)&16.432&7.385&10.384&8.783&11.897&11.644&9.795&10.653&$-$\\
MRE(\%)&9.752&-8.521&-0.331&-4.617&-1.384&-1.797&2.748&4.435&$-$\\
MARE(\%)&12.260&9.136&6.380&7.389&7.518&7.105&7.684&8.282&$-$\\
STDRE(\%)&12.115&7.225&7.746&8.287&9.425&9.048&9.945&9.986&$-$\\
\hline\hline
\end{tabular} 
\end{table*}

\begin{table*}
\label{latt-cons}
\caption{Band gaps using different functionals at their equilibrium lattice constant are
shown here. The mean absolute error is mentioned in the last row. All the experimental
references are collected from ref.~\cite{bgref}. The 
structures consider here are the same as it is given in Table-I.}
\begin{tabular}{c  c  c  c  c  c   c   c  c c}
\hline\hline
Solids&LSDA&PBE&PBEsol&TPSS&revTPSS&SCAN&TM-TPSS&TM&Expt.\\
\hline 
C&4.17 &4.13 &4.03 &4.17&4.04 &4.56 &4.15 &4.09 &5.48  \\
Si&0.46 &0.64 &0.48 &0.67 &0.57 &0.85 &0.65 &0.56 &1.17  \\
Ge&0.00 &0.00 &0.00 &0.00 &0.00&0.06 &0.23 &0.29 &0.74 \\
SiC&1.38 &1.47 &1.34 &1.42&1.30 &1.82 &1.47 &1.38 &2.42 \\
BN&4.48 &4.52 &4.36 &4.52 &4.38&5.04 &4.59 &4.49 &6.22 \\
BP&1.17 &1.28 &1.14 &1.29 &1.15&1.55 &1.28 &1.19 &2.4 \\
BAs&1.14 &1.22 &1.10 &1.21&1.10 &1.44 &1.19 &1.13 &1.46 \\
BSb&0.70 &0.75 &0.65 &0.65&0.54 &0.88 &0.61 &0.57 &n/a \\
AlP&1.47 &1.68 &1.50 &1.73&1.64 &1.95 &1.75 &1.63 &2.51 \\
AlAs&1.36 &1.54 &1.38 &1.59&1.51 &1.79 &1.59 &1.49 &2.23 \\
AlSb&1.11 &1.24 &1.13 &1.32 &1.23&1.39 &1.25 &1.16 &1.68 \\
$\beta-$GaN&1.82 &1.41&1.54 &1.31 &1.28 &2.05 &1.47 &1.48 &3.30 \\
GaP&1.45 &1.51 &1.52 &1.72 &1.60 &1.89&1.46 &1.56 &2.35 \\
GaAs&0.50 &0.15 &0.39 &0.38 &0.57&0.80 &0.80 &0.84 &1.52 \\
GaSb&0.11 &0.00 &0.00&0.00 &0.17 &0.12 &0.39 &0.46 &0.73 \\
InP&0.52 &0.37 &0.48 &0.54&0.59 &0.87 &0.71 &0.73 &1.42 \\
InAs&0.00 &0.00&0.00&0.00 &0.00 &0.00 &0.00 &0.00 &0.41 \\
InSb&0.00 &0.00&0.00&0.00 &0.00 &0.00 &0.00 &0.05 &0.23 \\
ZnS&1.89 &2.02 &2.09 &2.29&2.33 &2.71 &2.32 &2.31 &3.66 \\
ZnSe&1.28 &1.15 &1.24&1.45 &1.51 &1.80 &1.60 &1.59 &2.70 \\
ZnTe&1.31 &1.07 &1.24 &1.42&1.58 &1.62 &1.63 &1.64 &2.38 \\
CdS&0.97 &1.04 &1.01 &1.23&1.18 &1.47 &1.23 &1.21 &2.55 \\
CdSe&0.43 &0.49 &0.48 &0.71&0.70 &0.94 &0.81 &0.79 &1.90 \\
CdTe&0.66 &0.59 &0.93 &0.82&0.82 &0.97 &1.01 &1.00 &1.92 \\
MgO&5.13 &4.53 &4.68 &4.77 &4.72& 5.77&4.96 &4.88 &7.22 \\
MgS&3.14 &3.34 &3.35 &3.63 &3.64&4.19 &3.80 &3.70 &5.4 \\
MgSe&1.80 &1.84 &1.85 &2.14&2.16 &2.51 &2.23 &2.15 &2.47 \\
MgTe&2.41 &2.32 &2.35 &2.66&2.72 &3.03 &2.89 &2.74 &3.6 \\
CaS&2.00 &2.40 &2.18 &2.47 &2.46&2.84 &2.54 &2.43 &n/a \\
CaSe&1.73 &2.10 &1.90 &2.18&2.18 &2.55 &2.26 &2.15 &n/a \\
CaTe&1.33 &1.57 &1.37 &1.64&1.63 &2.14 &1.71 &1.80 &n/a \\
SrS&2.14 &2.52 &2.30 &2.59 &2.54&2.92 &2.57 &2.47 &n/a \\
SrSe&1.91 &2.25&2.05 &2.32 &2.29 &2.67 &2.33 &2.23 &n/a \\
SrTe&1.43 &2.09&2.12 &2.31 &2.38 &2.74 &2.51 &2.44 &n/a \\
BaS&1.14 &2.17 &1.98&2.26 &2.19 &2.52 &2.16 &2.08 &3.88 \\
BaSe&1.67 &1.97&1.79 &2.05 &2.01 &2.33 &2.01 &1.92 &3.58 \\
BaTe&1.30 &1.61&1.41 &1.67 &1.63 &1.94 &1.66 &1.56 &3.08 \\
\hline
MAE (eV)&1.158&1.154&{\bf\color{red}1.173}&1.033&1.055&{\bf\color{blue}0.736}&0.949&0.996&$-$\\
\hline\hline
\end{tabular} 
\end{table*}

\begin{table*}
\label{latt-cons}
\caption{Cohesive energies of 9 solids in eV/atom 
using different functionals at static-lattice constant are shown. The experimental
reference values are taken from ref.\cite{mgga-vasp}. The 
structures consider here are the same as it is given in Table-I.}
\begin{tabular}{c  c  c  c  c  c   c   c  c c}
\hline\hline
Solids&LSDA&PBE&PBEsol&TPSS&revTPSS&SCAN&TM-TPSS&TM&Expt.\\
\hline 
Li&1.786 & 1.583&1.653 &1.738 &1.625 &1.545 &1.664 &1.662 &1.658  \\
C&8.867 &7.714 &8.215 &7.420 &7.504 &7.899 &7.624 & 7.845& 7.545 \\
SiC&7.305 &6.356 &6.779 &6.298 &6.380&6.689 &6.478 &6.652 &6.478 \\
Si&5.194 &4.464 &4.810 &4.444&4.531 &4.811 &4.628 &4.788 &4.685 \\
LiF&4.867 &4.411 &4.515 &4.469 &4.389&4.784 &4.565 &4.554 &4.457 \\
LiCl&3.739 &3.332 &3.467 &6.442 &3.430&3.632 &3.551 &3.536 &3.586  \\
NaF&4.396 &3.962 &4.061 &4.272&3.944 &4.394 &4.163 &4.147 &3.970 \\
NaCl&3.438 &3.085 &3.197 &6.389&3.199 &3.438 &3.349 &3.326 &3.337 \\
MgO&5.982 &5.152 &5.441 &5.271&5.295 &5.654 &5.439 &5.496 &5.203 \\
\hline\hline
ME(eV/atom)&0.450&-0.163&0.068&-0.080&-0.139&0.147&-0.007&0.053&$-$\\
MAE(eV/atom)&{\bf\color{red}0.450}&0.201&0.153&0.152&0.139&0.171&{\bf\color{blue}0.041}&0.053&$-$\\
\hline
STDE(eV/atom)&0.394&0.153&0.258&0.162&0.077&0.137&0.067&0.109&$-$\\
MRE(\%)&8.593&-4.196&0.438&-0.952&-3.205&2.404&-0.136&0.769&$-$\\
MARE(\%)&8.593&4.696&2.670&3.496&3.205&3.905&0.837&0.769&$-$\\
STDRE(\%)&4.229&2.976&3.935&3.960&1.594&3.782&1.331&1.478&$-$\\
\hline\hline
\end{tabular} 
\end{table*}

We first perform the benchmark calculations for equilibrium lattice constant of
TM-TPSS and TM against the LSDA, PBE, PBEsol, TPSS, revTPSS and SCAN functional. For the
benchmark test set we consider 47 crystalline structures which include (i) semiconductor
diamond structures C, Si and Ge, (ii) zinc blende structures Sic, BN,
BP, BAs, BSb, AlP, AlAs, AlSb, $\beta-$GaN, GaP, GaAs, GaSb, InP, InAs, InSb, ZnS, ZnSe,
ZnTe, CdS, CdSe, CdTe, MgS, MgTe, (iii) ionic structures MgO, MgSe, CaS, CaSe, CaTe, SrS,
SrSe, SrTe, BaS, BaSe, BaTe, LiCl, LiF, NaCl, LiF, NaCl, NaF, and (iv) metal structures
Li, K, Al, Cu, Pd, Ag. All the calculations are performed using 
11$\times$11$\times$11 Monkhorst-Pack~\cite{monk} $\bf{k}$ meshes with tetrahedron
method~\cite{tetra}.
The self-consistence calulation of meta-GGAs (TPSS, revTPSS, SCAN, TM-TPSS and TM) are performed starting from
the converged 
wavefunction obtained in PBE calculations. 

\begin{table*}
\label{latt-cons}
\caption{Comparison between the calculated magnetic
moments ($M_s$/atom) and the experimental total magnetic moments of Fe, Co and Ni using
different functionals. Values are in $\mu$B. The Fe (bcc) and Ni (fcc) values of LSDA,
PBE, PBEsol, TPSS and revTPSS 
are taken from ref.~\cite{mgga-vasp}.}
\begin{tabular}{c  c  c  c  c  c   c   c  c c c c c c}
\hline\hline
Solids&Magnetic& &LSDA&PBE&PBEsol&TPSS&revTPSS&SCAN&TM-TPSS&TM&HSE06&Expt.\\
&Ordering&\\
\hline 
Fe (bcc)&FM&$a_0$ &2.747 &2.829 &2.782 &2.803 &2.794  &2.844 &2.811 &2.803&2.903 &2.853 (2.861)$^a$  \\
       &  &$M_s$ &1.97 &2.18 &2.11 &2.19 &2.20 &2.64 &2.25 &2.22 & 2.896&2.20$^a$\\

Co (fcc)&FM&$a_0$ &3.420 &3.514 &3.460 &3.479 &3.468 &3.479 &3.481 &3.469 & 3.547&3.537$-$3.558$^b$ \\
       &  &$M_s$ &1.49 &1.61 &1.57 &1.61 &1.62 &1.77 &1.64 &1.60 & 1.852&1.71$^c$ \\
    

Ni (fcc)&FM&$a_0$ &3.428 &3.520 &3.463 &3.481 & 3.465 &3.456 &3.468 &3.460 & 3.504&3.508 (3.516)$^a$ \\
       &  &$M_s$ &0.56 & 0.62 &0.60  & 0.63 &0.65 &0.66 &0.61 &0.60 & 0.839&0.61$^a$ \\
\hline\hline
\end{tabular} 
\footnotetext[1]{ref.~\cite{mgga-vasp}}
\footnotetext[2]{ref.~\cite{ferro}}
\footnotetext[3]{ref.~\cite{ferro2}}
\end{table*}

In Table-I, we list the benchmark calculations of TM-TPSS and TM with other semilocal
functionals. There we also calculate the mean (relative) error (ME/MRE), mean absolute
(relative) error (MAE/MARE)  and the standard deviation of the (relative) error
(STDE/STDRE). We obtain the maximum MAE using PBE and local density approximate (LSDA)
functional. The MAE of PBE and LSDA are obtained to be 0.078 \AA~ and 0.072 \AA~
respectively. LSDA has the tendency to underestimate the lattice constant while PBE
overestimates the lattice constant. The reduction in MAE is observed using meta-GGA
TPSS and revTPSS functionals. The TPSS functional performs slightly better than PBE  while
the revised version of TPSS (revTPSS) reduces the MAE significantly. Recently proposed
SCAN meta-GGA by Sun et. al.~\cite{SCAN15} gives
the same MAE as obtain from the PBEsol GGA functional. Interestingly, the lowest MAE is 
obtained from TM functional (with MAE 0.038 \AA~). The unmodified TPSS correction coupled
with TM exchange produce the equivalent MAE as obtain from PBEsol and SCAN. It is 
noteworthy to mention that the experimental lattice constants we consider hare not 
corrected for zero point anharmonic expansion (ZPAE) and all the experimental lattice
constant are taken from ref.~\cite{bgref,YMo16}. The results we obtain here using
the TM-TPSS and TM can be compared to the all electron calculation performed using
Gaussian09~\cite{gaussian} which is given in ref.~\cite{YMo16}. In ref.~\cite{YMo16} the
lowest mean absolute error is also reported using TM.  

\subsection{Bulk moduli}
The bulk modulus is defined as the change in the volume of the crystalline structures upon
acting the pressure. In terms of  total energy of the cell $E$, the bulk modulus is
expressed as $K = V\frac{\partial^2E}{\partial V^2}$. In density functional theory 
the bulk modulus is calculated at the equilibrium lattice constant $a_0$ by scanning over
the range of lattice constant (or volume). Several equation of states (EOS)~\cite{eos} are
available to fit the energy versus volume curve to obtain the bulk modulus. 
In our present case we used the post-processing code
VASPKIT~\cite{vaspkit}
to fit and obtain the bulk modulus of all the crystalline solids we compared here. The E
vs V output from VASP is used as an input of VASPKIT. The
VASPKIT is very well established post-processing code which is used to obtain several
post-processing calculations~\cite{vaspkit}. In VASPKIT, Murnaghan equation of 
states~\cite{eos} is used for the fitting. The test
set we use here and the performance of all the corresponding functionals are listed in
Table-II. For few selected solids we collect the values of LSDA, PBE, PBEsol, TPSS,
revTPSS from ref.~\cite{mgga-vasp}. For all other functionals and solids the
calculation is performed using the
same $\bf{k}$ points meshes as mentioned  in subsection-A.

From the results reported in Table-II it is indicative that the performance of TM-TPSS 
is best compared to all other functionals. The MAE of TM-TPSS even better than SCAN
functionals with MAE 7.195 GPa compared to MAE 7.785 GaP obtain from SCAN.  The
performance of SCAN, PBEsol and TPSS almost 
equivalent.  
It is also indicative that TM-TPSS performs better than TM functional though the
performance of TM is better than TM-TPSS in predicting lattice constant. The maximum MAE
is obtained from LSDA funcional. The performance of PBE is better than LSDA. The results  
we obtain using the TM functional are also very close to that obtain using G09 all
electron calculation reported by Mo et. al~\cite{YMo16}.

\subsection{Band gaps}

The band gap calculation using semilocal functional is fraught with difficulties due to
the absence of inherent non-locality and many electron self-interaction
(MESI)~\cite{Yangreview}. The 
hybrid functional proposed using semilocal exchange hole are very popular in predicting
the band gap for semiconductor materials accurately~\cite{HSE03,mpsk,jmhk,shk,bgref}. It
has been observed that the
meta-GGA functional implemented in generalized KS formalism give more realistic band gap
than GGA functionals~\cite{relbg}. Therefore, the improvement in band gap is observed
using meta-GGA type semilocal functionals compared to LSDA and GGA functionals. Here, we 
assess the performance of TM-TPSS and TM for 37 semiconductors at their equilibrium
lattice constant (reported in Table-I) of each functionals.

From Table-III, it is evident that all the density functionals underestimate the band
gap of all semiconductor which is obvious due to the absence of non-locality and MESI.
Interestingly, using SCAN meta-GGA we obtain more realistic band gap within 
all semilocal functionals. TPSS, revTPSS, TM-TPSS and TM perform equivalently in 
predicting band gaps. It is indicative from the obtained results of the SCAN, TM-TPSS and
TM that the SCAN functional outperformed the TM based functionals almost in every cases.
Interestingly, for Ge, InAs and InSb which are the difficult cases within semilocal
formalism, SCAN, TM-TPSS and TM perform extremely well and predict non-zero band gap except
only one for InAs. For Ge and InSb TM functional have non-zero band gap, whereas, for Ge 
both SCAN and TM based functionals predict non-zero values. Also, in all these cases the
TM functional performs better than SCAN functional. This is actually a most 
attractive feature of TM based functional than other semilocal functionals.

\subsection{Cohesive energies}
Cohesive energy is equivalent to the molecular atomization in the case of crystalline
solids. It is defined as the energy difference of the solid from its
neutral from as an atom. Finally the cohesive energy (in eV) per
atom is obtain by dividing the energy difference of the atoms in the unit cell. Here, we
consider a set of 9 crystalline solids to perform the benchmark calculations of all the
functionals. Among all the functionals under consideration TM-TPSS is accurate with
MAE of 0.041 eV/atom. In this case TM-TPSS is more accurate than TM functional. 
From Table-IV, we observed that the performance of TM-TPSS and TM are accurate 
compared to all other GGA and meta-GGA based functionals with MAE 0.041 eV/atom and 0.053 eV/atom
respectively. It is also noteworthy that TM has tendency to overestimate the cohesive
energy for all the crystalline solids considered here, whereas, TM-TPSS overestimates the
cohesive energies for few cases and underestimate for few cases. Overall, TM-TPSS performs
accurately in predicting cohesive energy of all the solids.

\subsection{Magnetic properties}
Studying strongly correlated systems within semilocal density functional is quite
difficult because of the different levels of interaction of $d$ and $f$ blocks. In Table-V
we calculate the magnetic moments and lattice constant of ferromagnetic Fe, Ni and Co.
Here  the results of all the semilocal functionals are also compared with the range
separated hybrid functional HSE06. All magnetic moments are calculated by optimizing the
structure with the corresponding semilocal functionals.  

The results presented in Table-V show that for Fe all the semilocal functional predicts
the lattice constant quite appropriately. In the case of magnetic moment, all the
functionals are accurate except LSDA and SCAN functional. For LSDA 
the underestimation of magnetic moment is observed. While SCAN 
overestimates the magnetic moment. In this case, both TM-TPSS and TM 
predict the magnetic moment accurately . In the case of Co, PBE is 
better than all other semilocal functional in predicting the lattice constant. LSDA
underestimate magnetic moment of Co, while SCAN is 
close to the experimental values. All other semilocal functional perform equivalently in
predicting magnetic moment. In case of Ni,
both the lattice constant and magnetic moment are obtained accurately within all the
semilocal functionals. Now we come to the discussion 
of range separated hybrid HSE06 in predicting all properties. HSE06 predicts accurately
various properties but computationally very expensive. In this case, HSE06 accurately
predict lattice constant for all the solids but overestimate the magnetic moments due to
the inclusion of too much HF exchange. This drawback has been discussed by Paier et.
al.~\cite{jmhk}.    


\section{Conclusions}
We assess the performance of TM functional in projector-augmented-wave
method with plane wave basis set for solid state band gaps, bulk moduli,
cohesive energies and magnetic properties. It has been shown that the
performance of TM-TPSS and TM are quite accurate within the popular
semilocal functionals in predicting all the properties except
semiconductor band gaps. In that case SCAN meta-GGA performs better than
TM. The TM-TPSS functional is accurate for bulk moduli and cohesive
energies, whereas, TM is accurate in predicting lattice constants. For
the band gap the performance of TM-TPSS is second best after SCAN. In
particular, TM-TPSS and TM predict non-zero band gap for several
semiconductor for which LSDA, PBE, PBEsol, TPSS and revTPSS predict
metallic. Lastly we conclude that the TM functional can be used
with confidence using plane wave basis in predicting all the solid state
properties accurately over the GGA and meta-GGA functionals.

\end{document}